\documentclass[12pt]{article}

\newcommand{\bbr}{I\!\! R}
\newcommand{\bbz}{Z\!\!\! Z}

\usepackage{graphicx} 

\begin{document}
\thispagestyle{empty}
\begin{center}

\null
\vskip-1truecm
\vskip2truecm
{\bf DE SITTER AND SCHWARZSCHILD-DE SITTER ACCORDING TO SCHWARZSCHILD AND DE SITTER\\}
\vskip2truecm
Brett McInnes
\vskip2truecm

Department of Mathematics, National University of Singapore, 10 Kent Ridge Crescent,
Singapore 119260, Republic of Singapore.\\ 
E-mail: matmcinn@nus.edu.sg\\    

\end{center}
\vskip1truecm
\centerline{ABSTRACT}
\baselineskip=15pt
\medskip
When de Sitter first introduced his celebrated spacetime, he claimed, following Schwarzschild, that its spatial sections have the topology of the real projective space $\bbr{P}^3$ (that is, the topology of the group manifold SO(3)) rather than, as is almost universally assumed today, that of the sphere $S^3$. (In modern language, Schwarzschild was disturbed by the non-local correlations enforced by $S^3$ geometry.) Thus, what we today call ``de Sitter space" would not have been accepted as such by de Sitter. There is no real basis within classical cosmology for preferring $S^3$ to $\bbr{P}^3$, but the general feeling appears to be that the distinction is in any case of little importance. We wish to argue that, in the light of current concerns about the nature of de Sitter space, this is a mistake. In particular, we argue that the difference between ``dS($S^3$)" and ``dS($\bbr{P}^3$)" may be very important in attacking the problem of understanding horizon entropies. In the approach to de Sitter entropy via Schwarzschild-de Sitter spacetime, we find that the apparently trivial difference between  $\bbr{P}^3$ and $S^3$ actually leads to very different perspectives on this major question of quantum cosmology. 
\vskip3.5truecm
\begin{center}

\end{center}

\newpage

\addtocounter{section}{1}
\section*{1. The Superfluous Antipodes}
Understanding de Sitter spacetime --- the finiteness of its entropy, its embedding in string theory --- is one of the key problems of current physics. Obviously, therefore, it is important to be confident that we know exactly what ``de Sitter spacetime" actually {\em is}. From its first appearance \cite{kn:desitter}, however, it has been clear that there is a fundamental ambiguity in the very definition of ``de Sitter spacetime", arising from the well-known fact that the Einstein equations do not completely fix the topology of spacetime. (de Sitter's paper can be found at $http://adsabs.harvard.edu/ads\_abstracts.html$.) This kind of topological ambiguity has recently attracted a great deal of attention from an {\em observational} point of view (see for example \cite{kn:levin}\cite{kn:weeks}), but far less attention has been paid to its theoretical implications. The purpose of this work is to draw attention to the fact that these theoretical implications have a bearing on issues of great current interest.

Today we are accustomed to the idea that the topology of de Sitter space is that of $\bbr \times S^3$, where $S^3$ is the three-sphere. (Sometimes it is convenient to use other slicings, such as that by flat three-dimensional spaces, but these do not cover the entire spacetime and so they tell us nothing about its global structure.) However, de Sitter himself explicitly rejected this interpretation: for him, the topology of ``de Sitter spacetime" was $\bbr \times \bbr{P}^3$. Here $\bbr{P}^3$ is the real projective space, obtained from $S^3$ by identifying all points with their antipodes. (The distinction between the two is precisely that between the group manifolds of SO(3) and SU(2); for it is easy to see that SU(2) has the topology of $S^3$, and it is well known that it covers SO(3) twice, so the group manifold of SO(3) is $\bbr{P}^3$.) The metric of $\bbr{P}^3$ is exactly that of $S^3$ --- that is, if the sectional curvature is $1/L^2$, it is
\begin{equation} 
g(\bbr{P}^3, 1/L^2) = g(S^3, 1/L^2) = L^2 [d\chi \otimes d\chi + sin^2(\chi)[ d\theta \otimes d\theta + sin^2(\theta)d\phi \otimes d\phi)]],
\end{equation}
but now the angles $\chi$, $\theta$, and $\phi$ (which on $S^3$ range respectively from zero to $\pi$, zero to $\pi$, and zero to $2\pi$) are subject to identifications according to the antipodal map $\aleph_3$ on $S^3$, defined by
\begin{eqnarray} 
\chi & \rightarrow & \pi - \chi \nonumber \\
\theta & \rightarrow & \pi - \theta \nonumber \\
\phi & \rightarrow & \pi + \phi.
\end{eqnarray}
Thus $\bbr{P}^3$ is the quotient $S^3/\bbz_2$ =  $S^3/\{ 1, \aleph_3 \}$.

Comparing the two, de Sitter states that $\bbr{P}^3$ ``is really the simpler case, and it is preferable to adopt this for the physical world." (He also reports a letter from Einstein to the effect that the latter agreed with him --- though it appears that Einstein later \cite{kn:luminet} changed his opinion, on extremely tenuous aesthetic grounds.) de Sitter was apparently influenced by a much earlier (1900) paper of Schwarzschild \cite{kn:schwarzschild1} (translated as \cite{kn:schwarzschild2}). Usually this paper is cited as one of the first attempts to discuss, from an observational point of view, the possibility that the spatial sections of the Universe may not have the geometry or topology of ordinary Euclidean space. Schwarzschild's paper is strange to modern eyes, however, in that, when he considers positively curved space, he {\em only} discusses $\bbr{P}^3$, which he calls ``the simplest of the spaces with spherical trigonometry." In fact, he explicitly rejects $S^3$ as a physically acceptable model for spatial geometry, on the grounds that the light emitted from a point in $S^3$ would collect again at the antipode, and ``one would not consider such complicated ({\em sic}) assumptions unless it were really necessary". Schwarzschild's point is that any two coplanar geodesics in $\bbr{P}^3$ intersect only once, while in $S^3$ they would do so twice. The intersection of geodesics is however a matter of local physics, and it is absurd that the geometry should try to enforce correlations on the largest possible length scales. When stated in this modern way, as a concern for the locality of physics, Schwarzschild's argument begins to seem very plausible; we shall further update it below. 

When, much later, the importance of isotropy and homogeneity came to be appreciated, it was realised by a few authors (see for example \cite{kn:hawking}, page 136) that even these stringent criteria provide {\em no} basis for discriminating against $\bbr{P}^3$ in favour of $S^3$. A precise mathematical discussion of the point can be given briefly as follows. Let $p$ be a point in a Riemannian manifold $M$, and let $X_p$ and $Y_p$ be arbitrary unit tangent vectors at $p$. Suppose that it is always possible to find an open set $U_p$ containing $p$ such that, when the metric on $M$ is restricted to $U_p$, there exists an isometry of $U_p$ mapping $X_p$ to $Y_p$. If this can be done for all $p$ in $M$ then the latter may be said to be {\em everywhere locally isotropic}. (If we insist that the isometries should be global, then of course one says that the manifold is globally isotropic; {\em but there is no observational warrant for this}.) In three space dimensions (though {\em not} in higher dimensions), an everywhere  locally isotropic manifold has to be a space of constant curvature; this follows from Schur's theorem (see \cite{kn:kobayashi} page 202). In the case of positive curvature, there are infinitely many such spaces, all of them being quotients of $S^3$ by finite groups. Even if we adjoin the condition of homogeneity (on Copernican grounds) the number of candidates remains infinite: for example, all of the lens spaces (see \cite{kn:wolf}) are homogeneous. (Curiously, this is a peculiarity of positive curvature: there are only finitely many everywhere locally isotropic, homogeneous three-manifolds of negative or zero curvature.) Thus, local isotropy and homogeneity do not single out either $S^3$ or $\bbr{P}^3$. But suppose that we extend the concept of homogeneity in the following way. Let us assume that, given any two pairs of points in $M$, (A, B) and (C, D), such that the distances d(A, B) and d(C, D) are equal, there exists an isometry of $M$ which maps A to C and B to D simultaneously. Then $M$ is said to be ``two-point homogeneous". This is a more restrictive form of homogeneity: instead of merely requiring all {\em points} in $M$ to be ``equally good", we are requiring this of {\em extended objects}. Now it can be shown \cite{kn:wolf} that the only positively curved examples in three dimensions are $S^3$ and $\bbr{P}^3$. Thus, even on the strictest interpretation of homogeneity, there are no grounds for preferring $S^3$. Nor is there any other basis for doing so: both are orientable spin manifolds, both have isometry groups of the maximal dimension (six), and so on.

The point of this historical/mathematical excursion is this. Schwarzschild and de Sitter were convinced that the topology of a positively curved world was that of $\bbr{P}^3$, and none of the relevant arguments from classical cosmology allow us to argue that they were wrong. In this sense, the $\bbr \times \bbr{P}^3$ version --- which we may call dS($\bbr{P}^3$) --- $\:${\em is}$\:$ the true ``de Sitter spacetime". {\em The fact that we tend to think otherwise today is merely a matter of historical accident.} 

The reader may object that while it is indeed true that $\bbr{P}^3$ is just as good as $S^3$ in classical cosmology, it does not follow that the same is true in quantum cosmology. One may wonder, for example, whether all is well with quantum field theory on dS($\bbr{P}^3$). This issue was addressed very comprehensively and precisely in \cite{kn:louko1}, where no difficulties were found; on the other hand, one of the more remarkable results of that work was the discovery that the response functions of (monopole) particle detectors were quite different in the two cases (dS($\bbr{P}^3$) and dS($S^3$)). Thus it is clear that even though ``the topology is behind the horizon", it has (in principle) measurable physical effects: out of sight is not out of mind. (We may mention here that there is no hope at present of distinguishing dS($\bbr{P}^3$) by direct astronomical observations, though this is not the case \cite{kn:weeks} for cosmologies with other positively curved locally isotropic homogeneous three-manifolds as spatial sections. The question as to whether the WMAP data \cite{kn:wmap} may possibly suggest that the spatial geometry or topology is non-trivial is currently under debate \cite{kn:efstathiou1}\cite{kn:efstathiou2}\cite{kn:tegmark}\cite{kn:banday}.)

More generally, however, dS($S^3$) is often favoured because it has a much larger isometry group than any other cosmological spacetime, {\em including} dS($\bbr{P}^3$). Even though $S^3$ and $\bbr{P}^3$ have (non-isomorphic) isometry groups of the same dimension, it is a remarkable fact that the isometry group of dS($\bbr{P}^3$) has the ``normal" size for an FRW cosmology --- it is six-dimensional --- whereas, as is notorious, dS($S^3$) has an isometry group of the largest possible dimension, namely ten. That distinction seems to favour dS($S^3$) as a ``background". Recently, however, it has been argued \cite{kn:goheer} very persuasively that ``the trouble with de Sitter space" is precisely that it has too many symmetries. These symmetries disrupt an attempt to understand de Sitter entropy in the context of one of the most promising approaches to horizon entropies (see \cite{kn:jacobson} for a recent general review), namely the ``horizon entropy as entanglement entropy" programme \cite{kn:israel}. {\em It seems that some symmetries of the de Sitter vacuum must be broken} if we are to understand de Sitter entropy. Since dS($\bbr{P}^3$) has a more ``normal" (FRW) isometry group than dS($S^3$), it is reasonable to ask whether dS($\bbr{P}^3$) can lead to a different approach.

The question of the real significance of de Sitter entropy is one of the deepest in physics, and we do not propose a complete answer here. The intention is much more modest: to show that dS($\bbr{P}^3$) does lead to a different perspective on the problem.  We argue that the apparently slight topological difference between the two actually has profound consequences for the entropy question, essentially because it enforces very different respective global structures for the Schwarzschild-de Sitter geometry (which has always played a crucial, if not very well-understood, role in deriving the formula for de Sitter entropy). In particular, we stress that the richer topology of the SdS($\bbr{P}^3$) conformal infinity (see Figure 5 below) offers better prospects for understanding de Sitter entropy as entanglement entropy. In fact, we argue that if we take the entanglement point of view to be fundamental, then the $\bbr{P}^3$ approach may explain why the Schwarzschild-de Sitter geometry is relevant to the computation of the entropy of pure de Sitter space. For, using recently developed global techniques for asymptotically de Sitter spacetimes, we show that Figure 5 gives a {\em generic} picture of the evolution of two entangled copies of $\bbr{P}^3$. No argument of this kind is possible in the $S^3$ case. 

In short, our main point is that, to understand de Sitter entropy, a choice must be made: dS($\bbr{P}^3$) or dS($S^3$). Our secondary purpose is to lay the foundations for an ``$\bbr{P}^3$ programme" by gathering together relevant data, particularly on Penrose diagrams and other aspects of the global structure of Schwarzschild-de Sitter space.

Before we proceed, we should clarify that we do {\em not} consider here the ``elliptic" version of de Sitter space, in which an antipodal identification is performed {\em both} in space and in time; here the identification will be purely spatial. Interesting discussions of the elliptic version (which was first suggested by F. Klein \cite{kn:kerszberg}) may be found in \cite{kn:schrodinger}\cite{kn:parikh}. Note that the fact that time reversal symmetry is broken rather strongly suggests (\cite{kn:visser}, page 288) that our universe is time-orientable. Elliptic de Sitter spacetime is not time-orientable, but dS($\bbr{P}^3$) is orientable both in time and in space. (For a further discussion of this aspect of elliptic de Sitter spacetime, see  \cite{kn:aguirre}.) We should also stress that our black holes are quite different from the superficially similar ``one-sided" black holes (see \cite{kn:louko2} for a review).

\addtocounter{section}{1} \section*{2. Basics of dS($\bbr{P}^3$)}

The global form of the de Sitter metric, corresponding to a positive cosmological constant $3/L^2$, is 

\begin{equation} 
g(dS_4) = - d\tau \otimes d\tau + L^2 cosh^2(\tau /L)[d\chi \otimes d\chi + sin^2(\chi)[ d\theta \otimes d\theta + sin^2(\theta)d\phi \otimes d\phi)]].
\end{equation}

The corresponding Penrose diagram, in the case of $S^3$ spatial sections, has the familiar form given in Figure 1. In this diagram, every point in the interior corresponds to a two-sphere, while those on the left and right vertical boundaries are the (of course, arbitrary) poles. The left and right-hand triangular regions are the two ``static patches", which may be described by coordinates (t, r), so that {\em both} poles correspond to r = 0. ( This is an important point, since it means that we must be careful if we wish to modify the geometry near ``r = 0". Note that there are of course infinitely many static patches: what we mean is that, given one static patch in this spacetime, one can always find another which does not intersect the first.) The diagonals are the cosmological horizons (corresponding to r = L).

\begin{figure}[!h]
\centering
\includegraphics[width=0.3\textwidth]{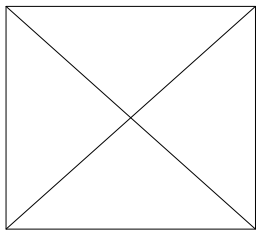}
\caption{Penrose diagram of dS($S^3$).}
\end{figure}

The metric in the ``static patches" is

\begin{equation} 
g(dS, static) = - (1 - {{r^2} \over {L^2}})dt \otimes dt \: + \:  (1 - {{r^2} \over {L^2}})^{-1}dr \otimes dr \: + \: r^2[ d\theta \otimes d\theta \: + \: sin^2(\theta)d\phi \otimes d\phi)].
\end{equation}
Note that the surfaces t = constant are represented in this diagram by curves which emanate from either pole and which all intersect at the centre of the diagram, that is, at the cosmological horizon.

Now consider $S^3$, and let $S^2_*$ be a distinguished ``equator". To construct $\bbr{P}^3$, note that every point in the Northern hemisphere is to be identified with its antipode in the Southern; hence we can discard the Northern hemisphere altogether, since its points have already been ``counted". This argument does not include $S^2_*$; along the equator, each $S^2$ point is identified with {\em its} antipode, producing exactly one copy of $\bbr{P}^2$, which must be included in $\bbr{P}^3$. Thus the reader can picture $\bbr{P}^3$ as a (three-dimensional) hemisphere with an $\bbr{P}^2$ equator. The Penrose diagram of dS($\bbr{P}^3$) therefore consists of three different kinds of vertical lines. First, at the left side of Figure 2, is just the worldline of the pole. Next, all interior points represent copies of $S^2$ ({\em not} $\bbr{P}^2$). Finally, at the extreme right-hand edge, the points (represented by stars) are copies of $\bbr{P}^2$. 
\begin{figure}[!h]
\centering
\includegraphics[width=0.4\textwidth]{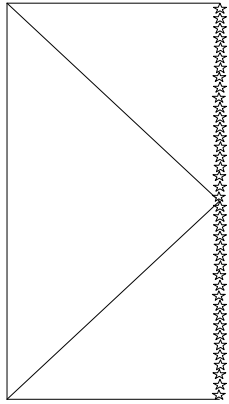}
\caption{Penrose diagram of dS($\bbr{P}^3$).}
\end{figure}
Now there is of course only one ``static patch" (that is, all static patches overlap and are in causal contact), which can be parametrised in the same way as before; notice that all of the t = constant sections terminate at a cosmological horizon which is a copy of $\bbr{P}^2$, and that the area of a surface r = constant is given by

\begin{eqnarray} 
A(r) = 4\pi r^2, \; \; r < L \nonumber \\
     = 2\pi r^2, \; \; r = L. \nonumber \\
\end{eqnarray}

 A ray of light moving radially outward from the pole will, if it leaves while the universe is still contracting (as reckoned by cosmic time $\tau$ in equation (3)), reach the $\bbr{P}^2$ surface and immediately re-appear on the opposite side of the sky (though this will never be seen by an observer at the pole, since the ray will ``reach" future infinity ``before" that --- one sees this in Figure 2 by letting the null geodesic ``bounce" off the right side of the diagram). Although $\bbr{P}^2$ is non-orientable, this gives rise to no problems for a three-dimensional object passing through it, because the object is not being parallel transported {\em within} $\bbr{P}^2$. It would return to the pole --- if that were possible --- with the directions perpendicular to its velocity vector reversed, but since, in three dimensions, there are two such directions, this constitutes a {\em rotation}, not a reflection. That is, an explorer who (of course, vainly)  attempts to circumnavigate dS($\bbr{P}^3$), setting out while the universe is still contracting, will see his home world turned upside-down --- that is, rotated through $\pi$ but not reflected --- after he passes through $\bbr{P}^2$.  In geometrical language, the linear holonomy group of $\bbr{P}^2$ is indeed the full group of rotations and reflections, O(2), but the linear holonomy group of $\bbr{P}^3$ is SO(3), which contains no reflections, only rotations. (Thus $\bbr{P}^3$ is an interesting example of a Riemannian manifold for which the holonomy group is connected, despite the fact that the manifold itself is not simply connected.) Note that these considerations show that an {\em odd}-dimensional ``de Sitter" spacetime with spatial sections of topology $\bbr{P}^{2n}$, $n \geq 1$, would be spatially non-orientable, which would conflict with the fact that parity is violated in our Universe (see for example \cite{kn:visser}, page 289). In a similar vein, $\bbr{P}^{1+4n}, \; n \geq 1$, while orientable, has a non-vanishing second Stiefel-Whitney class \cite{kn:lawson}, so it is not a spin manifold; hence dS($\bbr{P}^2$), dS($\bbr{P}^4$), dS($\bbr{P}^5$), and dS($\bbr{P}^6$) are all arguably unphysical in one way or another. It is pleasing that four-dimensional ``de Sitter spacetime" is singled out in this way.

The isometry group of $S^3$ is of course the orthogonal group O(4). The isometry group of $\bbr{P}^3$ may be found by recalling that the latter is precisely the Lie group manifold SO(3) with the usual Cartan-Killing metric. The isometries of SO(3) are given by left multiplication, right multiplication, and transposition of matrices. Recalling the formula $(ABC)^T = C^TB^TA^T$, we see that the isometry group consists of two copies of SO(3), plus a discrete symmetry which exchanges the two SO(3) factors. That is, the isometry group of $\bbr{P}^3$ is the six-dimensional group (SO(3) $\times$ SO(3)) $\triangleleft \; \bbz_2$, where the second product is semi-direct. This group is {\em not} isomorphic to O(4), though the Lie algebras are the same; in fact, since the universal covering group of SO(4) is given by two copies of SU(2), we have O(4)/$\bbz_2$ = (SO(3) $\times$ SO(3)) $\triangleleft \; \bbz_2$. Of course, this latter group is not a subgroup of O(4), any more than SO(3) is a subgroup of SU(2). Note that six is the maximum possible number of Killing vectors in three dimensions, so $\bbr{P}^3$ is, like $S^3$, maximally symmetric.

The isometry group of a ``normal" FRW cosmology is the same as the isometry group of its spatial sections (plus, perhaps, some finite factor). There is a good physical reason for this: FRW spacetimes represent time-dependent physical systems, so they should not have any continuous time symmetry. (A Big Bang/Crunch FRW spacetime has a discrete symmetry exchanging the Bang with the Crunch, but it has no {\em continuous} time symmetry.) As is well known, de Sitter space, in its dS($S^3$) form, is {\em not} a ``normal" FRW spacetime in this sense, for it has many other symmetries beyond O(4); in fact, it has an isometry group of the maximal possible size for a four-dimensional spacetime, the 10-dimensional group O(1,4). While this may be desirable for technical reasons, it is not necessary for quantum field theory (see \cite{kn:louko1}) and it is not a property of the real world --- some of the symmetries of the dS($S^3$) vacuum {\em must be broken}. Of course, one normally thinks of them as being broken by the introduction of matter and radiation, but the work of Goheer, Kleban, and Susskind \cite{kn:goheer} strongly suggests that the O(1,4) symmetry is broken in the vacuum, independently of the presence of any kind of matter (other than dark energy). 

The situation here is rather similar to the problem of gauge symmetry breaking in Calabi-Yau compactifications of string theory \cite{kn:polchinski}. There, gauge symmetries are broken by taking the Calabi-Yau space to have a non-trivial fundamental group (``Wilson loop symmetry breaking"), not by introducing matter fields with vacuum expectation values. In a similar way, simply by taking the spatial sections of de Sitter space to be $\bbr{P}^3$ instead of the simply connected sphere, we can break O(1,4), as follows.

dS($S^3$), with cosmological constant 3/$L^2$, is defined as the locus, in a five-dimensional flat space of the appropriate signature, given by
\begin{equation}
- A^2 + w^2 + x^2 + y^2 + z^2 = +L^2.   
\end{equation}
Simply by writing this as
\begin{equation}
w^2 + x^2 + y^2 + z^2 = A^2 + L^2,   
\end{equation}
we see that dS($S^3$) is globally foliated by copies of $S^3$, which are parametrised by the coordinates ($w,\;x,\;y,\;z$). The antipodal map on these spheres is given by the O(1,4) matrix diag($1, -1, -1, -1, -1$). Now clearly the only isometries of dS($S^3$) which descend to symmetries of dS($\bbr{P}^3$) are those which preserve the relationship of all antipodal pairs on $S^3$. That is, the only acceptable elements of O(1,4) are those which commute with the matrix diag($1, -1, -1, -1, -1$). These elements form a subgroup consisting of the matrix diag($-1, +1, +1, +1, +1$) together with the set of O(1,4) matrices of the form diag($1,\;$M), where M is any element of O(4). However, diag($1, -1, -1, -1, -1$) acts trivially on $\bbr{P}^3$, so we must project O(4) to O(4)/$\bbz_2$. Thus the full isometry group of dS($\bbr{P}^3$) is not the 10-dimensional group O(1,4) but rather the 6-dimensional group (with four connected components) $\bbz_2 \times$ [O(4)/$\bbz_2$], where the first $\bbz_2$ is generated by diag($-1, +1, +1, +1, +1$) and the second by diag($1, -1, -1, -1, -1$). Since it can be shown that the coordinate A is related to de Sitter proper time $\tau$ by A = L$\;$sinh($\tau$/L), and since we know that O(4)/$\bbz_2$ = (SO(3) $\times$ SO(3)) $\triangleleft \; \bbz_2$ (the isometry group of $\bbr{P}^3$), we see that the isometry group of dS($\bbr{P}^3$) is given simply by the isometry group of its spatial sections, together with a discrete symmetry exchanging past with future. (That is, the isometry group has been reduced from 10 dimensions to 6.) Thus, simply by taking the spatial sections to have a non-trivial fundamental group, we have succeeded in reducing the symmetries of ``de Sitter spacetime" to the standard form for a FRW cosmology. As in ``Wilson loop symmetry breaking" in string theory, this has been done without introducing matter of any kind. In the light of \cite{kn:goheer}, we take this as a strong hint that dS($\bbr{P}^3$) is the right version of de Sitter spacetime for our purposes.

If we now modify either dS($S^3$) or dS($\bbr{P}^3$) by introducing a black hole (necessarily of mass $M < L/(27)^{{1} \over {2}}$) around ``r = 0", then we obtain the Schwarzschild-de Sitter solution, with metric

\begin{equation} 
g(SdS) = - (1 - {{r^2} \over {L^2}} - {{2M} \over {r}})dt \otimes dt  \; + \;  (1 - {{r^2} \over {L^2}} - {{2M} \over {r}})^{-1}dr \otimes dr \; + \; r^2[ d\theta \otimes d\theta \; + \; sin^2(\theta)d\phi \otimes d\phi)].
\end{equation}
If we do this to dS($S^3$), then we are obliged to do it at both poles; but then we have no choice but to continue the process indefinitely, producing the Penrose diagram given in Figure 3 (see \cite{kn:gibhawk}). (We shall call this spacetime ``unwrapped" SdS($S^3$); the notation simply serves to remind the reader that we {\em began} with spatial topology $S^3$. The spatial topologies in Figures 3 and 4 are not, of course, the same as the topology of $S^3$.)
\begin{figure}[!h]
\centering
\includegraphics[width=0.5\textwidth,angle=90]{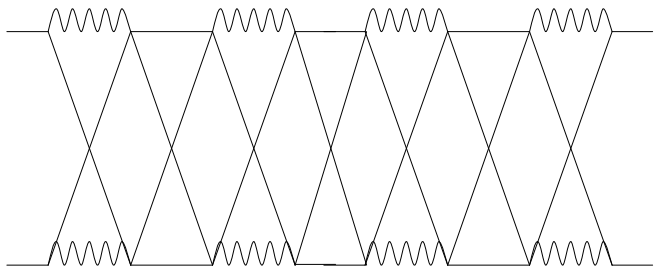}
\caption{Part of the (infinite) Penrose diagram of unwrapped SdS($S^3$)}
\end{figure}

Decorative though it may be, this represents a completely unphysical situation. It is true that such diagrams represent an idealised spacetime model, but it should be possible to understand the ways in which it would be necessary to modify them so as to obtain a more realistic model; that is, the idealised situation is physically interesting precisely to the extent that  it can be ``continuously deformed" to something more realistic. In the case of the maximally extended (Kruskal-Szekeres) Schwarzschild spacetime, for example, one understands that the formation of an actual black hole from a collapsing star would completely obscure the white hole region, leaving behind the physically relevant part of the diagram, and so on. The importance of this point for understanding the thermodynamics of black holes has been particularly emphasised, and explained very clearly, by Israel \cite{kn:israel}. But in the present case, one does not seriously suppose that the existence of a black hole in de Sitter spacetime necessarily entails the existence of another black hole at the other end of the Universe. Yet what else can we do with the infinite succession of superfluous black holes in Figure 3? This might be regarded as a modern, more pointed version of Schwarzschild's objection to $S^3$ as a model for physical space.
\begin{figure}[!h]
\centering
\includegraphics[width=0.3\textwidth]{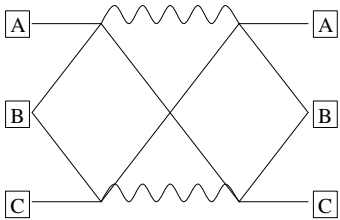}
\caption{Penrose diagram of wrapped SdS($S^3$)}
\end{figure}

The most obvious way to solve this problem is to notice that Figure 3 very much resembles the result of ``unwrapping" a cylinder. Reversing this ``unwrapping", we obtain ``wrapped" SdS($S^3$), pictured in Figure 4.
Here we understand that topological identifications are to be performed at the right and left edges, as indicated. (The topology of the spatial sections here is $S^1 \times S^2$; in Figure 3 it is $\bbr \times S^2$.) Henceforth, when we speak of ``SdS($S^3$)", we shall mean the spacetime pictured in Figure 4. 

We shall argue, however, that Figure 4 actually represents a situation scarcely less absurd than that in Figure 3. We now have a single black hole, it is true, but the ``Einstein-Rosen bridge" here actually leads back to our own universe. In the usual Schwarzschild case, the ``bridge" leads to another universe, which, like the white hole, would be eliminated in the process of formation of a real black hole; but here the ``other" universe which we so casually consign to non-existence {\em is in fact our own}. (If we blot out the white hole and the left side of Figure 4, replacing them as usual by the interior of a collapsing star, then what happens to the right side of the diagram at points A,B, and C?) We see that, unlike the Penrose diagram of the maximally extended Schwarzschild spacetime, {\em Figure 4 does not correspond to any physically realisable situation}; it is 
not an idealised representation of a real black hole in an asymptotically de Sitter spacetime. We shall return to this point below. (This problem does {\em not} arise for anti-de Sitter black holes, for reasons we shall explain in the next section.) 
\begin{figure}[!h]
\centering
\includegraphics[width=0.4\textwidth]{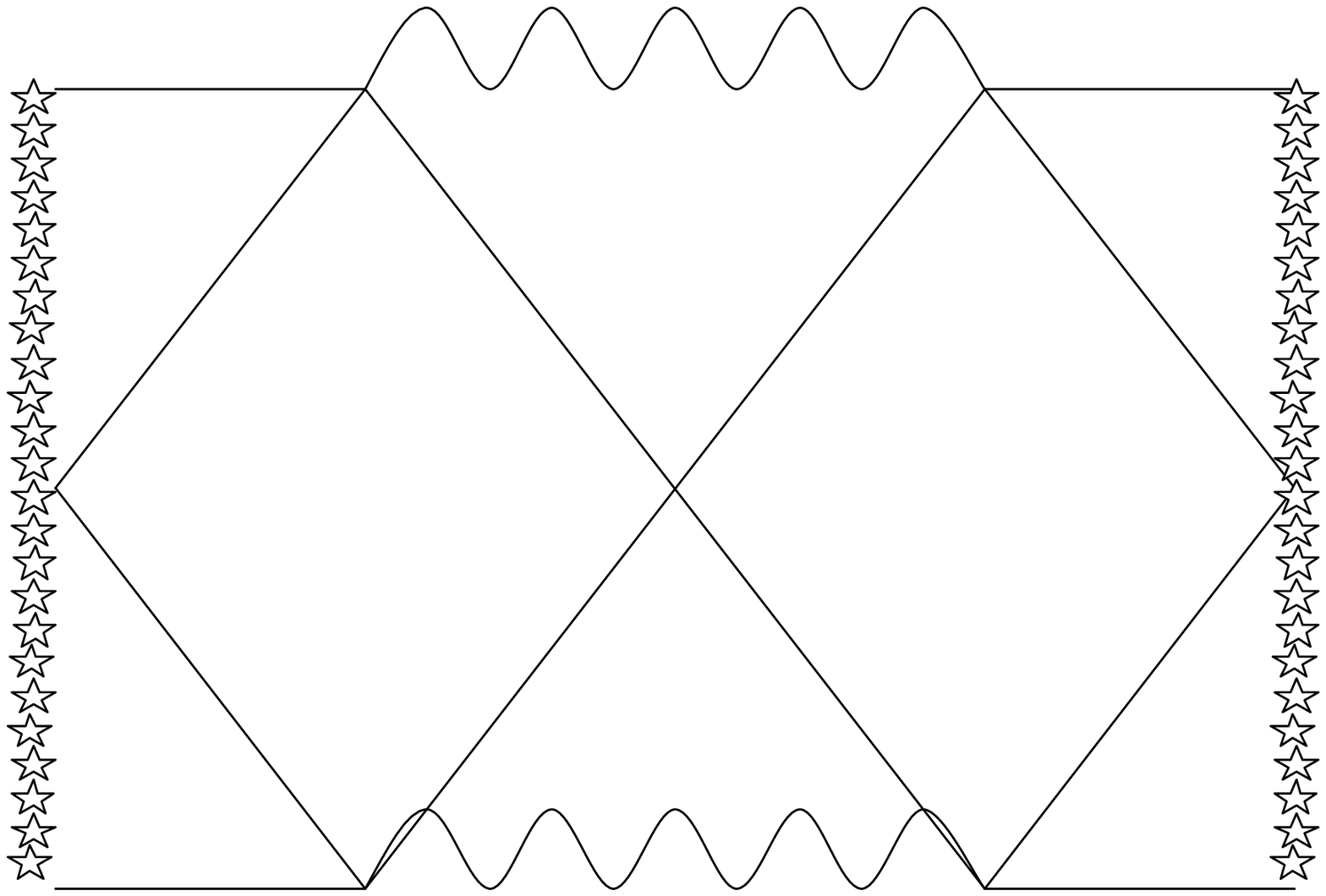}
\caption{Penrose diagram of SdS($\bbr{P}^3$)}
\end{figure}

If we now turn to dS($\bbr{P}^3$), and introduce a black hole (with $M < L/(27)^{{1} \over {2}}$) into it, we find that these absurdities are eliminated in the most natural way. 
Now there is only one point with r = 0, so the introduction of a black hole there has no consequences for the antipode, simply because there {\em is} no antipode. The maximal extension need have no other singularities, and the Schwarzschild-de Sitter metric now describes a perfectly reasonable spacetime which we shall call SdS($\bbr{P}^3$). (Again, this notation simply reminds the reader of the general context --- $\bbr{P}^3$ as opposed to $S^3$. The spatial topology will be discussed below: it is certainly not the same as that of $\bbr{P}^3$.) The Penrose diagram is pictured in Figure 5.
Here we see the ``worldlines" of two {\em distinct} copies of $\bbr{P}^2$ at left and right. (No topological identification of the two lines of stars is intended.) The diagram can be interpreted in just the same way as the Penrose diagram of the maximally extended Schwarzschild spacetime: there is a white hole, an Einstein-Rosen bridge leading to another, different universe, and so on. The formation of a black hole from the collapse of a star is represented in the usual way. In view of the fact that both Schwarzschild and de Sitter believed in antipodal identification, one is doubly tempted to claim that Figure 5 represents the ``true" Schwarzschild-de Sitter spacetime geometry.

Now let us turn to the application of these ideas to the question of horizon entropy. 

\addtocounter{section}{1}
\section*{3. Horizon Entropy: dS($S^3$) vs dS($\bbr{P}^3$) }

In this section we shall argue that the spacetime represented by Figure 5 is particularly interesting from the point of view of entanglement entropy \cite{kn:israel}. The reason is that, in a sense to be explained, Figure 5 is what we obtain, {\em generically}, when we take two copies of $\bbr{P}^3$, join them by a ``bridge", and allow them to become entangled. The point is that the general structure of Figure 5, in which the two sides remain causally disconnected, is obtained independently of symmetries and initial conditions. Before discussing that, let us clarify some more elementary points.

A celebrated result of Gibbons and Hawking \cite{kn:gibhawk} states that there is an entropy associated with the horizon of de Sitter space, given by one quarter of the horizon area. The argument, both in its original form and in the subsequent simplified presentations, has several mysterious aspects, so we shall present it here very briefly, following Bousso \cite{kn:bousso}. This will also allow us to clarify the relevance of the fact that the horizon of dS($\bbr{P}^3$) with a given cosmological constant has precisely half the area of the corresponding dS($S^3$) horizon.

We begin not with de Sitter spacetime but rather with the Schwarzschild-de Sitter spacetime discussed above. We take it that the black hole is small ($M << L/(27)^{{1} \over {2}}$), with energy E = M. There are two horizons, the black hole horizon with radius $r_+$, and the cosmological horizon with radius $r_{++}$; we have
\begin{equation} 
1 \; - \; {{2E} \over {r_{++}}} \; - \; {{r^2_{++}} \over {L^2}} = 0,
\end{equation}
or
\begin{equation} 
E = {{1} \over {2}}r_{++}[1 \; - \; {{r^2_{++}} \over {L^2}}].
\end{equation}
Thus we may think of $r_{++}$ as varying with E:
\begin{equation} 
{{dE} \over {dr_{++}}} = - 1 \; + \; {{3} \over {2}}[1 \; - \; {{r^2_{++}} \over {L^2}}].
\end{equation}
The area of the horizon in dS($\bbr{P}^3$) is of course given by A = $2\pi r^2_{++}$. Assuming as usual that the entropy is proportional to the horizon area, $S = \zeta A$, we obtain 
\begin{equation} 
{{dS} \over {dE}} = {{4\pi \zeta r_{++}} \over {- 1 \; + \; {{3} \over {2}}[1 \; - \; {{r^2_{++}} \over {L^2}}]}}.
\end{equation}

Now by considering a system consisting of initially well-separated matter and a black hole, Bousso argues that we should take $dE = -dM$. The first law of thermodynamics now gives us
\begin{equation} 
T = {{{{1 \; - \; {{3} \over {2}}[1 \; - \; {{r^2_{++}} \over {L^2}}]}}} \over {{4\pi \zeta r_{++}}}}
\end{equation}
for the temperature of the de Sitter horizon. For a very small black hole, $r_{++}$ is approximately equal to L (the de Sitter horizon radius), and so we have, approximately, 
\begin{equation} 
T = {{1} \over {4\pi \zeta L}}.
\end{equation}
Now temperature is a strictly local quantity, proportional to the surface gravity \cite{kn:wald}. The surface gravities for the cosmological horizons of dS($S^3$) and dS($\bbr{P}^3$) are the same --- only the areas differ, {\em not} the radii. If we had done the above calculation in dS($S^3$), we would have obtained a temperature of $1/8\pi \zeta L$. Since the two answers must agree, the constant $\zeta$ differs in the two cases:   
\begin{equation} 
\zeta_{\bbr{P}^3} = 2\zeta_{S^3}.
\end{equation}

Thus for example if the entropy is one quarter of the horizon area in dS($S^3$), then it is half the horizon area in dS($\bbr{P}^3$) --- which is the same numerical value in each case, namely $\pi L^2$. Thus, there is no discrepancy in the value of the entropy in the two cases, only in the way in which that value is computed. Similarly, the ``cosmological entropy bound" \cite{kn:bousso} still reads
\begin{equation} 
S_{matter} \; \leq \; \pi R_gR_c,
\end{equation}
where $S_{matter}$ is the entropy of some matter system in an asymptotically de Sitter spacetime, $R_g$ is the corresponding Schwarzschild radius, and $R_c$ is the radius of the cosmological horizon; the various factors of 1/2 cancel, and so one still has agreement between equation (14) and the cosmological version of the well-known Bekenstein bound, which requires in general that $S_{matter} \leq 2\pi ER$, where $E$ ( = $R_g/2$) is the energy of the system and $R$ is its ``largest dimension" ( = $R_c$, which has the same value in dS($\bbr{P}^3$) as in dS($S^3$)).

All this is perhaps less surprising if we examine equations (5), for it is clear that the areas of the surfaces r = constant are given by the conventional formula for any such surface which lies inside, {\em but arbitrarily close to,} the horizon. Thus, we should not expect the two entropies to  differ unless we think of the entropy as being {\em exactly} localised on the horizon. (If we computed the area by taking a surface with a radius infinitesimally less than that of the horizon, then we would find no difference at all.) In short, there is no profound difference between dS($\bbr{P}^3$) and dS($S^3$) as far as the {\em magnitude} of the entropy is concerned. It is otherwise, however, when we ask how this entropy should be {\em understood}.

Perhaps the strangest aspect of the above derivation --- this comment applies to both dS($\bbr{P}^3$) and dS($S^3$) --- is the fact that we had to introduce a black hole into otherwise singularity-free ``de Sitter spacetime". As it stands, this is a mere device: the black hole is introduced, used in the calculation, and then removed to obtain the result. It is natural to suspect, however, that something deeper is afoot. One suspects that the black hole will play some fundamental role in obtaining a deeper understanding of de Sitter entropy.

Black hole entropy is itself notoriously difficult to understand, but there is a growing consensus that {\em black hole entropy is entanglement entropy} $\;$ \cite{kn:uglum}\cite{kn:callan}. (See \cite{kn:israel} for a review.) The idea, in general, is to try to understand the thermodynamics of a quantum system by taking two, {\em independent} copies of the system, and then coupling them in a particular way so as to obtain a pure state. The quantum entanglement of the two systems then implies that each sub-system acts as a ``heat bath" for the other --- tracing over the modes hidden to one side results in an apparently thermal density matrix for that side. It is of course very natural to apply this picture to the (maximally extended) Schwarzschild spacetime, with the two systems corresponding to the two distinct universes on each side of the Einstein-Rosen bridge. The standard proportionality between the black hole entropy and its horizon area can be obtained in this way. (There are technical complications arising from field-theoretic divergences; see \cite{kn:chang} for recent developments.) Similar ideas have recently been used in several very deep investigations of eternal asymptotically anti-de Sitter black holes (see particularly \cite{kn:maldacena}, \cite{kn:ooguri}, and \cite{kn:hubeny}).

The Penrose diagram for an anti-de Sitter black hole (see Figure 6 --- note that it is not square, see \cite{kn:hubeny}) is similar to that of the extended Schwarzschild black hole, in the sense that it has an Einstein-Rosen bridge leading to another universe; thus it is natural to think of the thermal properties of the black hole as arising from the entanglement of the universes on each side of the bridge. 
\begin{figure}[!h]
\centering
\includegraphics[width=0.3\textwidth]{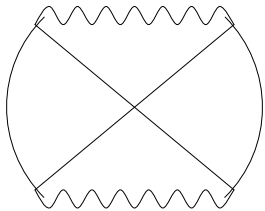}
\caption{Penrose diagram of eternal AdS black hole}
\end{figure}
{\em Because the conformal boundary of this spacetime consists of two disconnected components}(the curved lines at left and right in Figure 6), one  is led very naturally to suppose that this entanglement is encoded in an entanglement of two independent conformal field theories on the two boundaries. This in turn leads to profound insights into, for example, the black hole information paradox \cite{kn:maldacena}. 

In \cite{kn:goheer} it was suggested that the entropy of the horizon of dS($S^3$) might be understood in a similar way by means of entanglement between the two static patches. Apparently intractable problems arise in that approach from the excessively ``large" symmetry group of dS($S^3$). While we saw above that dS($\bbr{P}^3$) solves this problem in a very elegant way, dS($\bbr{P}^3$) also does away with the ``second" static patch. In view of the fact that the standard derivations of the de Sitter entropy formula involve, as we saw above, the use of the Schwarzschild-de Sitter spacetime, one might think that a more natural approach would be to try to follow the example of the Schwarzschild and anti-de Sitter cases and attempt to couple two independent systems on either side of the Einstein-Rosen bridge in either SdS($S^3$) or SdS($\bbr{P}^3$).

For SdS($S^3$) there are, however, two serious objections to this plan. Firstly, we argued earlier that it is far from clear that SdS($S^3$) bears any relation to the real world. Secondly, SdS($S^3$) does not have a bridge connecting two {\em independent} universes: it has instead a ``handle" leading from one universe back to itself. (See Figure 4.) Thus, the situation in SdS($S^3$) is not analogous to the Schwarzschild and anti-de Sitter black hole spacetimes --- we do not really have ``two systems" as we did in those cases. One might try to argue that the two static patches appearing in Figure 4 are causally disconnected {\em from each other}, and therefore constitute two ``independent" systems, but this will only work if we are allowed to ignore the fact that these systems {\em have exactly the same conformal infinities} (future and past; recall the topological identification of the left and right edges of Figure 4). A more detailed explanation of the difference between the anti-de Sitter black hole and its $S^3$ de Sitter counterpart runs as follows. As is clear from Figure 1, de Sitter space has a conformal boundary which consists of two components, one each in the future and past. For simplicity, let us concentrate on the future boundary. Evidently it is connected. Now the conformal boundary of anti-de Sitter space is also connected, but, as we saw, the conformal boundary of the anti-de Sitter black hole is {\em disconnected}. This splitting of the boundary in the anti-de Sitter case is quite crucial for Maldacena's analysis of entanglement in that case \cite{kn:maldacena}. But now examine Figure 4: clearly, the future boundary is {\em connected} in the $S^3$ de Sitter case, whether or not a black hole is present. (In other words, the number of boundary components does not double). Even if one does not believe in a ``dS/CFT correspondence" \cite{kn:strominger}, it is unreasonable to go to the other extreme and assume that the conformal boundary has no influence whatever on the de Sitter bulk. (See \cite{kn:staruszkiewicz} for a discussion of the intimate relation of bulk and boundary in de Sitter space, a relationship which exists quite independently of any one-to-one ``correspondence"; see also \cite{kn:schaar} for an answer to the concerns raised in \cite{kn:susskind}.) So it seems clear that the entanglement approach to horizon entropies is unlikely to work here.

If indeed we do not really have two independent systems on the two sides of the SdS($S^3$) black hole, then the entanglement programme cannot even get off the ground. It seems that the differing global structures of the anti-de Sitter and de Sitter black holes explain the fact that the thermal properties of de Sitter space are so much more difficult to understand. What underlies this global difference, and how are we to apply the entanglement programme to de Sitter spacetime?

The answer is to be found by taking the ``entanglement argument" to be {\em fundamental}, in the following sense. Consider a  spacetime of the kind which has identifiable spatial sections --- such spacetimes need not be globally hyperbolic, as the important example of anti-de Sitter spacetime shows. (Anti-de Sitter spacetime, in global coordinates, has spatial sections with the topology of $\bbr^3$). Suppose that the spatial sections have the topology of a three-dimensional manifold, $M_3$. Now temporarily ignore the original spacetime and let us consider the problem of understanding the thermodynamics associated with the geometry or physical contents of $M_3$. In order to use entanglement to do this, we take {\em two} copies of $M_3$ and join them by a ``bridge". The idea is to regard the two copies of $M_3$ as containing independent systems, which are to become entangled via the bridge. We then propose to use the Einstein equations, in the usual initial-value formulation, to construct a full four-dimensional spacetime having this new three-manifold as its spatial sections. This spacetime can then be used to understand the thermodynamics of the original spacetime. Of course, the construction may not be a straightforward matter: there will be technical questions such as the dependence on the choice of the metric and extrinsic curvature on the three-manifold, and potential problems associated with non-globally hyperbolic spacetimes (such as the one pictured in Figure 6). In particular, {\em we have no guarantee that the spatial sections will evolve in a way that preserves the independence of the two systems}, since we do not know, in general, how to ensure that conformal infinity will split, as in Figure 6. But leaving these difficulties to one side for the present, let us try to understand the basic conditions required for the idea to work.

The model here, of course, is anti-de Sitter space. As mentioned above, the spatial sections here are $\bbr^3$. To construct a bridge between two copies of $\bbr^3$, we remove a ball from each copy, and then paste in a cylinder to close the resulting edges. This process is called ``taking the connected sum" of the two copies of $\bbr^3$ (see \cite{kn:besse} for interesting applications of this idea in differential geometry). It is easy to see that, by doing this, we obtain a space which is very different from $\bbr^3$: in the usual topological notation for the connected sum,
\begin{equation} 
\bbr^3 \# \bbr^3 \neq \bbr^3.
\end{equation}
If we take the manifold $\bbr^3 \# \bbr^3$ and allow it to evolve in accordance with the Einstein equations (with a negative cosmological constant as source), then, as expected, various complications arise; for example, we will need to construct an analytic extension of the domain of dependence of the initial spacelike hypersurface, and it is not clear that the two sides of the bridge will {\em necessarily} remain independent (as we require if we want to use entanglement). But at least the basic necessary condition (17) is satisfied: there {\em are} indeed two sides to the bridge. In fact, of course, we know that it is at least {\em possible} to begin with $\bbr^3 \# \bbr^3$ and generate a full asymptotically anti-de Sitter spacetime in which the two sides of the bridge remain out of causal contact (yet entangled): for that is just a description of the spacetime pictured in Figure 6. In other words, we can think of the anti-de Sitter black hole, which has  
$\bbr^3 \# \bbr^3$ as its spatial sections, as arising from an effort to construct a three-dimensional space consisting of two copies of $\bbr^3$, causally disconnected but entangled quantum-mechanically via a bridge.
 
If we now wish to replicate this approach in the de Sitter case, we must consider taking two copies of $S^3$ and constructing the connected sum. But it is easy to see that
\begin{equation} 
S^3 \# S^3 \; = \; S^3,
\end{equation}
and so the most basic necessary condition fails to be satisfied. It is simply not possible to join two {\em different} copies of $S^3$ with a bridge and obtain something new. The best we can do is to join $S^3$ to itself with a ``handle", {\em and this is precisely why we were driven to the structure for SdS($S^3$) given by Figure 4.} But we have argued that this will not answer: we need a bridge to another universe, not one which leads back to our own. The simple topological facts represented by (17) and (18)  underlie the fact that SdS($S^3$) apparently cannot be used to explicate de Sitter entropy by means of quantum entanglement.

As we have stressed, the $\bbr{P}^3$ version of Schwarzschild-de Sitter space, SdS($\bbr{P}^3$) (Figure 5), is very different to SdS($S^3$). The spatial sections of this spacetime are obtained by taking two independent copies of $\bbr{P}^3$, deleting a ball from each (think of the ball as the black hole event horizon and its interior), and pasting in a cylinder --- in short, the spatial sections are copies of $\bbr{P}^3 \# \bbr{P}^3$. Let us therefore try to repeat the above argument, starting now with $\bbr{P}^3$ instead of $S^3$. Immediately we find that there is a fundamental topological  difference between SdS($\bbr{P}^3$) and SdS($S^3$), for we have 
\begin{equation} 
\bbr{P}^3 \# \bbr{P}^3 \; \neq \; \bbr{P}^3.
\end{equation}
(The proof will be given below.) Thus, from the point of view of the entanglement programme, Figure 5 represents a situation more similar to Figure 6 than to Figure 4; and that is what we need if we are to use Schwarzschild-de Sitter geometry to understand de Sitter entropy. It is clear that the two systems on either side of the SdS($\bbr{P}^3$) bridge are indeed, on any reckoning, independent: in particular, we see that the presence of the black hole splits the conformal boundary, just as it does in anti-de Sitter spacetime. (That is, the full boundary now has {\em four} connected components, instead of the two in Figure 2.) It therefore seems at least possible that SdS($\bbr{P}^3$) can shed some light on the question of thermality in de Sitter spacetime, via the entanglement programme.

In fact, in the remainder of this section we shall argue that SdS($\bbr{P}^3$) is exceptionally well-suited to the entanglement approach. As we mentioned above, we cannot in general expect that the two sides of a ``bridge" will {\em necessarily} remain independent, as they do in Figures 5 and 6, when a connected sum of three-manifolds evolves from initial data; this could easily depend on symmetries, the choice of initial data, and so on. Indeed, as we shall see, it is actually possible for $\bbr{P}^3 \# \bbr{P}^3$ to evolve in such a way that future conformal infinity is connected, as it is in Figure 4, so that the two sides of the bridge do {\em not} remain independent. However, one might hope that this behaviour is not {\em generic} in the special case of asymptotically de Sitter spacetimes, where we expect the time evolution to be well-behaved. (That is, we do not expect difficulties with global hyperbolicity, as in the anti-de Sitter case.) Ideally, we would like to show that the time evolution of generic $\bbr{P}^3 \# \bbr{P}^3$ initial data {\em inevitably} leads to the formation of a black hole (an $\bbr{P}^3$ Schwarzschild-de Sitter black hole in the spherically symmetric case), for that would ensure that the general geometry pictured in Figure 5 is typical. If this is true, with a reasonable definition of ``generic", then we will have a better understanding of the fact that Schwarzschild-de Sitter geometry plays a key role in the analysis of de Sitter entropy. 

Actually, using techniques recently developed by Andersson and Galloway \cite{kn:galloway}, and assuming a form of cosmic censorship, we can show that a statement of this kind is indeed true, generically. To see the need for the ``generic" restriction, and to understand how to use the Andersson-Galloway techniques, we need a more detailed description of the topology of $\bbr{P}^3 \# \bbr{P}^3$, which is of considerable interest in its own right.

The principal point for the reader to grasp in what follows is that taking a connected sum of a manifold with itself {\em can} --- but need not --- drastically modify the structure of the fundamental group. For example, the connected sum of two manifolds with finite fundamental groups is a space with a fundamental group which is finite in some cases (such as $\bbr^3 \# \bbr^3$ and  $S^3 \# S^3$, which are simply connected), but infinite in others. $\bbr{P}^3$ has the simplest possible non-trivial fundamental group, namely $\bbz_2$, the group with only two elements (since passing to the antipode twice brings us back to the starting point). It does not follow from this, however, that the fundamental group of $\bbr{P}^3 \# \bbr{P}^3$ is finite. In fact it is not. One can prove this in various ways, but it will be most enlightening to exhibit the fundamental group explicitly. 

Consider the line $\bbr$, regarded as a Riemannian manifold, and let $R_0$ and $R_{\pi}$ be the isometries
\begin{equation} 
R_0 \; : \; x \; \rightarrow \; -x
\end{equation}
\begin{equation} 
R_{\pi} \; : \; x \; \rightarrow \; 2\pi - x,
\end{equation}
where $x \in \bbr$. These are reflections in 0 and $\pi$ respectively. Obviously both are of finite order, but the composite $R_{\pi}\circ R_0$ is just translation by $2\pi$, and so $R_0$ and $R_{\pi}$ generate an infinite discrete group. Now let $\aleph_{2}$ be the antipodal map on $S^2$, and define two isometries of $\bbr \times S^2$ by
\begin{equation} 
A_0 \; : \; (x, \; s) \; \rightarrow \; (R_0(x), \; \aleph_2 (s)) 
\end{equation}
\begin{equation} 
A_{\pi} \; : \; (x, \; s) \; \rightarrow \; (R_{\pi}(x), \; \aleph_2 (s)), 
\end{equation}
where $s \in S^2$. Since $R_0$, $R_{\pi}$, and $\aleph_2$ are all orientation-reversing on the spaces on which they act, $A_0$ and $A_{\pi}$ both preserve orientation; and since $\aleph_{2}$ has no fixed point, the complicated infinite, non-abelian group generated by $A_0$ and $A_{\pi}$ acts freely on $\bbr \times S^2$. Call this group $\Gamma$; then [$\bbr \times S^2$]/$\Gamma$ is a compact orientable three-dimensional manifold with a metric which descends from $\bbr \times S^2$ (since $\Gamma$ acts isometrically) and with $\bbr \times S^2$ as its universal (Riemannian) cover. As such, it is one of the well-known ``locally homogeneous" three-manifolds which arise in the Thurston programme; see \cite{kn:scott}. Now $A_{\pi}\circ A_0$ is just translation by $2\pi$ on $\bbr$ and the identity on $S^2$, so $\Gamma$ contains the infinite cyclic group $\bbz$. Recalling that the circle is just $\bbr /\bbz$, one can use this to show that [$\bbr \times S^2$]/$\Gamma$ is a principal fibre bundle, with structure group U(1), over $\bbr{P}^2$. Since [$\bbr \times S^2$]/$\Gamma$ is orientable and $\bbr{P}^2$ is not, it follows that the bundle is non-trivial. 

Now return to Figure 5 and consider the following construction. Take a point on the right-hand cosmological horizon (remember that it is a copy of $\bbr{P}^2$) and send out a spacelike geodesic, perpendicular to the horizon. The geodesic enters the black hole horizon, passes through the bridge, emerges from the black hole horizon on the other side, and reaches the {\em other} horizon (on the left of the diagram). It then emerges from the other side of the sky in that universe, and passes through the event horizon of the left-hand black hole on the opposite side of its event horizon --- note that, unlike the cosmological horizon, the black hole horizon has the topology of $S^2$, so this is not the same point as the one from which it emerged. The spacelike geodesic then continues back through the tunnel; it pierces each $S^2$ it encounters on the opposite side from the point it passed through on the first leg of its journey. It emerges back into the original universe on the opposite side of the right-hand black hole horizon from the point where it entered. It then continues out to the cosmological horizon and re-appears on the opposite side of the sky --- at its starting point: so the geodesic we have described is in fact a (topological) {\em circle.} By varying the starting point on $\bbr{P}^2$, we can clearly cause such a circular geodesic to pass through any given point of $\bbr{P}^3 \# \bbr{P}^3$. Locally, $\bbr{P}^3 \# \bbr{P}^3$ is therefore the product of a circle with $\bbr{P}^2$, though obviously this is not so globally: in fact, $\bbr{P}^3 \# \bbr{P}^3$ is just a non-trivial principal U(1) bundle over $\bbr{P}^2$. Since there is (up to bundle isomorphism) a unique such bundle over $\bbr{P}^2$ (see \cite{kn:scott}), we have
\begin{equation} 
\bbr{P}^3 \# \bbr{P}^3\; = \;  [\bbr \times S^2]/\Gamma.
\end{equation}
This way of thinking about $\bbr{P}^3 \# \bbr{P}^3$ will be useful to us in various ways. The first point to note is that this equation implies that the universal cover of $\bbr{P}^3 \# \bbr{P}^3$ is $\bbr \times S^2$, and this establishes (19) above, since the universal cover of $\bbr{P}^3$ is of course $S^3$. Secondly, we see at once that the fundamental group of $\bbr{P}^3 \# \bbr{P}^3$ is the {\em infinite} non-abelian group $\Gamma$, despite the fact that the fundamental group of $\bbr{P}^3$ is just $\bbz_2$. 

We can now proceed with our discussion of spacetimes evolving from $\bbr{P}^3 \# \bbr{P}^3$ in accordance with the Einstein equations (with a positive cosmological constant term as source).

First, we note that it is {\em not} the case that such a spacetime must {\em always} evolve into a black hole spacetime. To see this, recall that $\bbr{P}^3 \# \bbr{P}^3$ can be regarded as a principal U(1) fibre bundle over $\bbr{P}^2$. Therefore each point in $\bbr{P}^3 \# \bbr{P}^3$ projects to a point in $\bbr{P}^2$ which is contained in an open set such that $\bbr{P}^3 \# \bbr{P}^3$ is locally trivial over that open set. It follows that $\bbr{P}^3 \# \bbr{P}^3$ can be described locally by coordinates $\theta, \; \phi$ (the usual polar coordinates on $S^2$, projected to $\bbr{P}^2$) together with an angular coordinate $\psi$ (running from 0 to $2\pi$) describing position on any circular fibre. (Here it may be useful to examine Figure 5: $\psi$ = 0 corresponds to the base $\bbr{P}^2$, and $\psi$ = $\pi$ at the ``other" $\bbr{P}^2$. Values of $\psi$ between $\pi$ and $2\pi$ cannot be seen in Figure 5 because of the suppression of the other angular coordinates; but it is important to remember that the spacelike geodesic we described above passes through each $S^2$ on the opposite side on the way back to its starting point, so values of $\psi$ between $\pi$ and $2\pi$  refer to different points to values of $\psi$ between 0 and $\pi$.) Using these coordinates, one can define a Lorentzian metric on the four-dimensional manifold $\bbr \times (\bbr{P}^3 \# \bbr{P}^3)$ by 
\begin{equation} 
g(N) = - d\tau \otimes d\tau + {{L^2} \over {3}} cosh^2(\sqrt{3}\tau /L)d\psi \otimes d\psi + {{L^2} \over {3}}[d\theta \otimes d\theta + sin^2(\theta)d\phi \otimes d\phi)].
\end{equation}
This spacetime is globally hyperbolic, with $\bbr{P}^3 \# \bbr{P}^3$ as its spatial sections; like the de Sitter metric g(dS) (equation (3)) the metric g(N) satisfies the Einstein equations in the form 
\begin{equation} 
Ricci(g(N)) \; = \; {{3} \over {L^2}}\;g(N),
\end{equation}
and the spacetime is of course entirely non-singular --- actually it is a local ({\em not} global) product of two-dimensional de Sitter space with $\bbr{P}^2$. Evidently it is not true that $\bbr{P}^3 \# \bbr{P}^3$ necessarily evolves into a black hole spacetime.

This spacetime is in fact the $\bbr{P}^3$ version of the Nariai spacetime (see \cite{kn:bousso}); let us call it N($\bbr{P}^3$). Recall that the Nariai metric is obtained as the limit of the Schwarzschild-de Sitter metric in which the two horizons coincide, that is, when the mass of the black hole satisfies $M = L/(27)^{{1} \over {2}}$. Thus we are now considering SdS($\bbr{P}^3$) (Figure 5) in that limit. 
\begin{figure}[!h]
\centering
\includegraphics[width=0.4\textwidth]{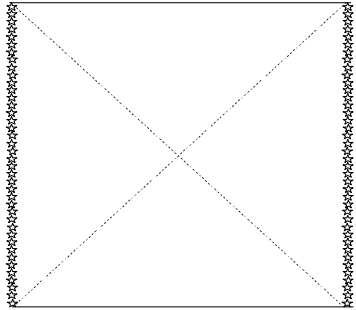}
\caption{Penrose diagram of N($\bbr{P}^3$).}
\end{figure}     
The Penrose diagram of N($\bbr{P}^3$) is given in Figure 7. It is square because the coordinate $\psi$ runs from 0 to $\pi$ from right to left, and the conformal time coordinate $\eta$ defined by sin($\eta$) = sech($\sqrt{3}\tau /L$) runs from 0 to $\pi$ from bottom to top. (The stars, as before, indicate a suppressed $\bbr{P}^2$.) The reader is asked however to bear in mind that $\psi$ runs from 0 to $2\pi$; it may be helpful to imagine the diagram as being drawn on the surface of a cylinder. (The diagonal lines are dotted so as to remind us of this.) There is a further subtlety                                                                                                                                                                                                                                                                                                                                                                                                    here which cannot be represented in the diagram: if we write g(N) in its conformal form,
\begin{equation} 
g(N) = {{L^2} \over {3sin^2(\eta)}} [ - d\eta \otimes d\eta + d\psi \otimes d\psi + sin^2(\eta)[d\theta \otimes d\theta + sin^2(\theta)d\phi \otimes d\phi)]],
\end{equation}
then we see at once that the conformal metric (obtained by removing the factor $L^2/3sin^2(\eta)$) assigns zero area to all  points at $\eta$ = 0 and $\pi$. Thus the horizontal lines in Figure 7 have a different meaning to those in Figures 2 and 5; they are similar to the {\em vertical} lines in Figures 1 and 2, they represent {\em points}. That is, the suppressed copies of $S^2$ and $\bbr{P}^2$ shrink to zero size at the top and bottom of the diagram. From the conformal point of view,  the Nariai spacetime contains a bridge which collapses in finite conformal time; thus we see the sense in which the Nariai spacetime is the ``limit" of the Schwarzschild-de Sitter spacetime (Figure 5) as $M \; \rightarrow \; L/(27)^{{1} \over {2}}$, since the latter contains a bridge which (partly) collapses in finite conformal (but also {\em proper}) time. 

It is clear from Figure 7 that the conformal future of N($\bbr{P}^3$) is connected, just as in Figure 4; and so, as in that case, N($\bbr{P}^3$) fails to maintain the independence of the two sides of the $\bbr{P}^3 \# \bbr{P}^3$ bridge. However, it is also clear from our detailed construction that N($\bbr{P}^3$) is anything but {\em generic}. We saw, firstly, that it is obtained from the Schwarzschild-de Sitter spacetime SdS($\bbr{P}^3$) by means of an exact fine-tuning of the mass of the original black hole: the slightest deviation of the mass M from $L/(27)^{{1} \over {2}}$ will either return us to Figure 5 or produce a naked singularity (if $M \; > \; L/(27)^{{1} \over {2}}$). Secondly, the expansion of the spacelike surfaces in N($\bbr{P}^3$) is also extremely fine-tuned, in the sense that the $\theta$ and $\phi$ directions do not expand at all (equation (25)), while of course the $\psi$ direction expands ever more rapidly. A generic, physically reasonable solution of equation (26) should expand in all directions, though not necessarily to the same extent. This brings us to the work of Andersson and Galloway \cite{kn:galloway}.

Following \cite{kn:galloway}, we say that a spacetime $M_4$ has a {\em regular future conformal completion} if $M_4$ can be  regarded as the interior of a spacetime-with-boundary $X_4$, with a ({\em non-degenerate}) metric $g_X$ such that the boundary is {\em spacelike} and lies to the future of all points in $M_4$, while $g_X$ is conformal to $g_M$, that is, $g_X = \Omega^2g_M$, where $\Omega = 0$ along the boundary but $d\Omega \neq 0$ there. A spacetime with a regular past conformal completion is then defined in the obvious way. In physical language, spacetimes of this kind are ``generically de Sitter-like", that is, no directions are ``left behind" by the cosmic expansion. For example, the Nariai spacetime  does {\em not} have a regular future or past conformal completion; as we saw, it cannot be called ``generic". 

A spacetime with a regular future (past) conformal completion is said to be {\em future (past) asymptotically simple} if every future (past) inextendible null geodesic has an endpoint on future (past) conformal infinity. In physical language, one would say that such spacetimes are free of singularities in either the future or the past. Figure 5 represents an example of a spacetime which is globally hyperbolic and has regular future and past conformal completions, but which is neither future nor past asymptotically simple, while of course Figure 2 pictures a spacetime which is both future and past asymptotically simple.

With this background, we can now state the very remarkable and powerful theorem of Andersson and Galloway (\cite{kn:galloway}, Remark 4.1). Suppose that a spacetime satisfies $Ricci(X,\; X) \geq 0$ for any null vector X, is globally hyperbolic, and has regular future and past conformal completions. {\em If the fundamental group of the Cauchy surfaces of the spacetime is  infinite}, then the spacetime can be neither future nor past asymptotically simple.

Of course, any spacetime satisfying equation (26) satisfies $Ricci(X,\; X) \; = \; 0$ for any null vector X. Unlike anti-de Sitter space, de Sitter space is globally hyperbolic, so it is natural to impose global hyperbolicity here: we can think of it as the version of Cosmic Censorship (see \cite{kn:wald}) which is suitable in this context. (Cosmic censorship can be expected to hold when, as in spacetimes satisfying equation (26), the dominant energy condition is valid; again, it may (therefore) very well fail in asymptotically anti-de Sitter spacetimes; see \cite{kn:maeda}. We would have to be careful about this point if we introduced (say) a dilaton field into an asymptotically de Sitter spacetime.) As we have just discussed, the requirement that the spacetime should have regular future and past conformal completions is just a kind of ``genericity condition" which eliminates special cases like Nariai spacetime. Finally, as we saw earlier, $\bbr{P}^3 \# \bbr{P}^3$ has an infinite fundamental group. Hence, the Andersson-Galloway theorem allows us to conclude, {\em independently of any specific assumptions about symmetry or initial conditions,} that a physically reasonable, generic evolution of an $\bbr{P}^3$ bridge produces a spacetime which is neither future nor past asymptotically simple. That is, like SdS($\bbr{P}^3$), the spacetime {\em must} have (spacelike) singularities to the future and the past, which, as in Figure 5, split future and past conformal infinity and ensure the independence of the two systems on either side of the bridge. {\em Hence the claim that the $\bbr{P}^3$ version of de Sitter spacetime is better suited to the entanglement approach than the $S^3$ version.} 

Note that the crucial condition here, because it is the condition not satisfied by the $\bbr{P}^3$ spatial sections of dS($\bbr{P}^3$), is the fact that $\bbr{P}^3 \# \bbr{P}^3$ has an infinite fundamental group. Thus it is the {\em topology} of the bridge that ensures the independence of the systems on either side of it. As always, topological arguments have the virtue of being immune to perturbations or symmetry violations.

Further light is shed on the role of the black hole in de Sitter entropy if we consider black hole evaporation, which should of course be highly relevant here. (The fact that de Sitter black holes do always evaporate is not obvious; see \cite{kn:bousso} for a discussion.) 

The evaporation of an $\bbr{P}^3$ black hole is pictured, under the usual assumption that there is no remnant, in Figure 8.
\begin{figure}[!h]
\centering
\includegraphics[width=0.4\textwidth]{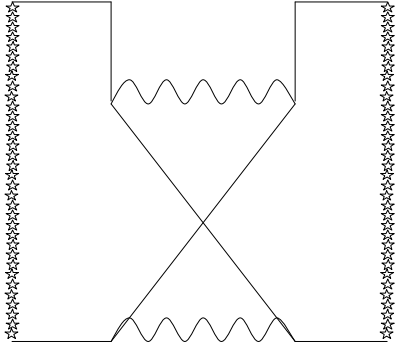}
\caption{Penrose diagram of the evaporation of SdS($\bbr{P}^3$).}
\end{figure}
As can be seen, the spacetime ``near" future conformal infinity in this case corresponds to a pair of copies of the upper half of Figure 2; that is, the evaporation of a black hole in dS($\bbr{P}^3$) just gives us two independent copies of the later history of dS($\bbr{P}^3$). (As usual, the stars indicate copies of a suppressed $\bbr{P}^2$.) In fact, the situation here is quite similar to the evaporation of a maximally extended Schwarzschild black hole, which also produces two disconnected universes. Once again, however, the contrast with the situation in dS($S^3$) is sharp, for the evaporation of a black hole like the one pictured in Figure 4 actually maintains the connectedness of future infinity --- the Penrose diagram will resemble the upper half of Figure 1, and the topology of future infinity will be that of {\em one} copy of $S^3$. (As Bousso explains \cite{kn:bousso}, the evaporation of a black hole corresponds to the ``pinching off" of $S^1 \times S^2$, which does not disconnect the space; instead it produces a copy of $S^3$, so that the later history of the spacetime resembles that of dS($S^3$). In fact one can think of the two ``r = 0" points in Figure 1 --- the left and right vertical lines --- as the relics of the pinching point. In the same way, the two vertical lines to either side of the upper singularity in Figure 8 correspond to ``r = 0", but they do not represent the same point; they are separate relics of the pinching. In this case, however, the pinching disconnects the spatial sections.) Thus dS($\bbr{P}^3$) ``proliferates" by means of the evaporation of an ordinary black hole, while dS($S^3$) does  not. (This proliferation is quite different to the kind discussed in \cite{kn:bousso2}, which arises from {\em multiple} singularities nucleating in a pre-existing Nariai spacetime. Further discussions of that kind of proliferation may be found in \cite{kn:nojiri1}\cite{kn:nojiri2}; see also \cite{kn:medved} for another perspective on the Nariai spacetime.)

For our purposes Figure 8 is not very different from Figure 5, and nothing we have said about the quantum entanglement of the two sides of the bridge needs to be modified by the fact that the black hole evaporates. The novelty here is that an observer at an event in, say, the top right hand side of the diagram will not observe a black hole, for it has by then evaporated; yet we can still think about the entanglement of a system on the right side of the diagram with another one on the left. Again we see that the black hole is essential to the understanding of thermodynamics even in (what appears to be) pure de Sitter space. (Notice in this connection that the Hawking radiation will not re-focus; it will be ``held back" by the accelerated expansion of the universe, for circumnavigations of the universe are not possible in dS($\bbr{P}^3$), as Figure 2 shows, unless they begin in the infinite past. Thus the Hawking radiation is eternally available for inspection.)

The upshot is that we can conclude that SdS($\bbr{P}^3$) and its evaporating version, portrayed in Figures 5 and 8, provide the right environment for discussing entanglement entropy in the de Sitter context. At the very least, we now understand the otherwise mysterious fact that a black hole is relevant to de Sitter entropy.

We close this section with some remarks on relations to other work.

The importance of entanglement between the {\em boundary components} of de Sitter space has been particularly emphasised by Balasubramanian, de Boer, and Minic \cite{kn:minic}, who argue that (ordinary) de Sitter space is holographically dual to a {\em pair} of entangled conformal field theories defined respectively on future and past conformal infinity. Here we will have many other interesting variants: if we call the {\em four} components of conformal infinity in Figure 5 the ``left future", ``left past", ``right future", and ``right past", then we can consider entanglements of the conformal field theory on the ``right future" with either the conformal field theory on the ``right past" or one on the ``left past", and so on. We believe that this will lead to a better understanding of both the black hole and the cosmological entropies. A suitable generalisation of the techniques of reference \cite{kn:das} should be useful here; see also \cite{kn:schaar}. The fact that the Schwarzschild-de Sitter spacetime has two horizons and therefore two ``temperatures" is of course one of its most notable features, and has been the subject of many interesting recent studies: see for example \cite{kn:shank}\cite{kn:teitelboim}. We expect to relate this to the multiplicity of ``infinities" in our view of this spacetime. 

Another interesting extension would be to extend Maldacena's arguments \cite{kn:maldacena}\cite{kn:hubeny} from Figure 6 to Figures 5 and 8, in the hope of elucidating the nature of the Schwarzschild-de Sitter singularity. In the anti-de Sitter case, a crucial role is played by correlators connecting the two disconnected components of conformal infinity. Again, there are more possibilities here for ``left-future/right-future" correlators, and so on. 

{\em None} of these possible ways of exploring Schwarzschild-de Sitter spacetime would even make sense if we were to try to  use the model of that spacetime pictured in Figure 4. Whether or not one accepts the claim that Figure 5 (or Figure 8) is the ``right" model, there can now be no doubt that the distinction between the $S^3$ and $\bbr{P}^3$ versions of de Sitter spacetime is radical. And this is the point we wished to make. 

\addtocounter{section}{1}
\section*{4. Conclusion }
In this work we have explained the profound differences between $\bbr{P}^3$ de Sitter spacetime and its more familiar $S^3$ counterpart, with a view to applications to horizon thermodynamics. Ultimately these differences stem from two simple mathematical facts: first, the symmetry group of dS($\bbr{P}^3$) is much smaller than that of dS($S^3$); and second, $\bbr{P}^3 \# \bbr{P}^3$ differs very drastically from $\bbr{P}^3$, while $S^3 \# S^3$ is precisely the same as $S^3$. It is reasonable to expect that the $\bbr{P}^3$ approach will at the very least lead to new insights into the nature of horizon entropy.

Hawking, Maldacena, and Strominger \cite{kn:HMS} have proposed a completely different approach to understanding de Sitter entropy in terms of quantum entanglement, one which does not involve Schwarzschild-de Sitter spacetime. Unfortunately it is difficult to extend their approach to four dimensions, but the idea --- which involves the study of a de Sitter braneworld in anti-de Sitter space --- deserves further attention. In particular, one can ask what happens in that approach when ``de Sitter spacetime" is interpreted as dS($\bbr{P}^3$). We shall return to this question elsewhere.

Bousso \cite{kn:bousso} has argued that, while charged black holes may not be directly physical, they need to be taken into account in the study of horizon entropy. It is clear that a charged black hole in dS($\bbr{P}^3$) will have many interesting properties. In fact, such black holes require the use of techniques from ``Alice physics", introduced by Krauss, Wilczek, and Preskill \cite{kn:krauss}\cite{kn:preskill} (see \cite{kn:mcinnes} for cosmological applications), and they are currently under investigation from that point of view.

\section*{Acknowledgement}
The author is profoundly grateful to Soon Wanmei, for producing the drawings, for several valuable discussions, and for marrying him while this work was being done.


\begin{thebibliography}{18}
\linespread{0.5}
\bibitem{kn:desitter}
W. de Sitter, A. Einstein's Theory of Gravitation and its Astronomical Consequences, Third Paper, MNRAS 78 (1917) 3.
\bibitem{kn:levin}
Janna Levin, Topology and the Cosmic Microwave Background, Phys.Rept. 365 (2002) 251, arXiv:gr-qc/0108043 
\bibitem{kn:weeks}
Jean-Philippe Uzan, Alain Riazuelo, Roland Lehoucq, Jeffrey Weeks, Cosmic microwave background constraints on multi-connected spherical spaces, arXiv:astro-ph/0303580 
\bibitem{kn:luminet}
Jean-Pierre Luminet, Past and Future of Cosmic Topology, Proceedings of ``Concepts de l'Espace en Physique", Les Houches, 1997, arXiv:gr-qc/9804006 
\bibitem{kn:schwarzschild1}
K Schwarzschild, \"Uber das zul\"assige Kr\"ummungsmass des Raumes, Vierteljahrschrift d. Astronom. Gesellschaft 35 (1900) 337
\bibitem{kn:schwarzschild2}
K. Schwarzschild, On the permissible curvature of space, Class. Quantum Grav. 15 (1998) 2539
\bibitem{kn:hawking}
S.W. Hawking and G.F.R. Ellis, The Large Scale Structure of Spacetime, Cambridge University Press, 1973
\bibitem{kn:kobayashi}
S. Kobayashi, K. Nomizu {\em Foundations of Differential Geometry I}, Interscience, 1963
\bibitem{kn:wolf}
J.A. Wolf  {\em Spaces of Constant Curvature}, Publish or Perish, 1984
\bibitem{kn:louko1}
Jorma Louko, Kristin Schleich, The exponential law: Monopole detectors, Bogoliubov transformations, and the thermal nature of the Euclidean vacuum in $\bbr{P}^3$ de Sitter spacetime, Class.Quant.Grav. 16 (1999) 2005, arXiv:gr-qc/9812056 
\bibitem{kn:wmap}
C. Bennett et al, First Year Wilkinson Microwave Anisotropy Probe (WMAP) Observations: Foreground Emission, arXiv:astro-ph/0302208 
\bibitem{kn:efstathiou1}
G. Efstathiou, Is the Low CMB Quadrupole a Signature of Spatial Curvature?, arXiv:astro-ph/0303127 
\bibitem{kn:efstathiou2}
G. Efstathiou, The Statistical Significance of the Low CMB Mulitipoles, arXiv:astro-ph/0306431 
\bibitem{kn:tegmark}
Angelica de Oliveira-Costa, Max Tegmark, Matias Zaldarriaga, Andrew Hamilton, The significance of the largest scale CMB fluctuations in WMAP, arXiv:astro-ph/0307282
\bibitem{kn:banday}
H. K. Eriksen, F. K. Hansen, A. J. Banday, K. M. Gorski, P. B. Lilje, Asymmetries in the CMB anisotropy field, arXiv:astro-ph/0307507 
\bibitem{kn:goheer}
Naureen Goheer, Matthew Kleban, Leonard Susskind, The Trouble with de Sitter Space, arXiv:hep-th/0212209 
\bibitem{kn:jacobson}
Ted Jacobson, Renaud Parentani, Horizon Entropy, Found.Phys. 33 (2003) 323, arXiv:gr-qc/0302099 
\bibitem{kn:israel}
W. Israel, Black Hole Thermodynamics, in {\em Current Trends in Relativistic Astrophysics}, Eds L. Fernandez-Jambrina and L.M. Gonzalez-Romero, Springer, 2003
\bibitem{kn:kerszberg}
P. Kerszberg, {\em The invented universe : the Einstein-De Sitter controversy
(1916-17) and the rise of relativistic cosmology}, Oxford University Press, 1989
\bibitem{kn:schrodinger}
E. Schr\"odinger, {\em Expanding Universes,} Cambridge University Press, 1956
\bibitem{kn:parikh}
Maulik K. Parikh, Ivo Savonije, Erik Verlinde, Elliptic de Sitter Space: $dS/\bbz_2$, Phys.Rev. D67 (2003) 064005, arXiv:hep-th/0209120 
\bibitem{kn:visser}
M. Visser, {\em Lorentzian Wormholes from Einstein to Hawking,} AIP Press, 1995 
\bibitem{kn:aguirre}
Anthony Aguirre, Steven Gratton, Inflation without a beginning: a null boundary proposal, Phys.Rev. D67 (2003) 083515, arXiv:gr-qc/0301042 
\bibitem{kn:louko2}
Jorma Louko, Single-exterior black holes, Lect.Notes Phys. 541 (2000) 188, arXiv:gr-qc/9906031
\bibitem{kn:lawson}
H. B. Lawson, M.L. Michelsohn, {\em Spin Geometry,} Princeton University Press, 1989
\bibitem{kn:polchinski}
J. Polchinski, {\em String Theory}, Cambridge University Press, 1998  
\bibitem{kn:gibhawk}
G.W. Gibbons, S.W. Hawking, Cosmological Event Horizons, Thermodynamics, and Particle Creation, Phys.Rev.D15 (1977), 2738
\bibitem{kn:bousso}
Raphael Bousso, Adventures in de Sitter space, arXiv:hep-th/0205177 
\bibitem{kn:wald}
R.M. Wald, {\em General Relativity}, Chicago University Press, 1984
\bibitem{kn:uglum}
L. Susskind, J. Uglum, Black Hole Entropy in Canonical Quantum Gravity and Superstring Theory, Phys.Rev. D50 (1994) 2700, arXiv:hep-th/9401070 
\bibitem{kn:callan}
Curtis Callan, Frank Wilczek, On Geometric Entropy, Phys.Lett. B333 (1994) 55, arXiv:hep-th/9401072 
\bibitem{kn:chang}
Darwin Chang, Chong-Sun Chu, Feng-Li Lin, Transplanckian Entanglement Entropy, arXiv:hep-th/0306055 
\bibitem{kn:maldacena}
Juan M. Maldacena, Eternal Black Holes in AdS, JHEP 0304 (2003) 021, arXiv:hep-th/0106112 
\bibitem{kn:ooguri}
Per Kraus, Hirosi Ooguri, Stephen Shenker, Inside the Horizon with AdS/CFT, Phys.Rev. D67 (2003) 124022, arXiv:hep-th/0212277 
\bibitem{kn:hubeny}
Lukasz Fidkowski, Veronika Hubeny, Matthew Kleban, Stephen Shenker, The Black Hole Singularity in AdS/CFT, arXiv:hep-th/0306170 
\bibitem{kn:strominger}
A. Strominger, The dS/CFT Correspondence, JHEP 0110 (2001) 034, arXiv:hep-th/0106113 
\bibitem{kn:staruszkiewicz}
Aharon Casher, Pawel O. Mazur, Andrzej J. Staruszkiewicz, De Sitter Invariant Vacuum States, Vertex Operators, and Conformal Field Theory Correlators, arXiv:hep-th/0301023 
\bibitem{kn:schaar}
Jan Pieter van der Schaar, Inflationary Perturbations from Deformed CFT, arXiv:hep-th/0307271 
\bibitem{kn:susskind}
Lisa Dyson, James Lindesay, Leonard Susskind, Is There Really a de Sitter/CFT Duality?, JHEP 0208 (2002) 045, arXiv:hep-th/0202163 
\bibitem{kn:besse}
A.L. Besse, {\em Einstein Manifolds}, Springer, 1987
\bibitem{kn:galloway}
L. Andersson, G. J. Galloway, dS/CFT and spacetime topology, arXiv:hep-th/0202161
\bibitem{kn:scott}
P. Scott, The Geometries of Three-Manifolds, Bull.Lond.Math.Soc. 15 (1983) 401 
\bibitem{kn:maeda}
Thomas Hertog, Gary T. Horowitz, Kengo Maeda, Generic Cosmic Censorship Violation in anti de Sitter Space, arXiv:gr-qc/0307102 
\bibitem{kn:bousso2}
Raphael Bousso, Proliferation of de Sitter Space, Phys.Rev. D58 (1998) 083511, arXiv:hep-th/9805081
\bibitem{kn:nojiri1}
S. Nojiri, S.D. Odintsov, Quantum evolution of Schwarzschild-de Sitter (Nariai) black holes, Phys.Rev. D59 (1999) 044026, arXiv:hep-th/9804033
\bibitem{kn:nojiri2}
Andrei A. Bytsenko, S. Nojiri, S.D. Odintsov, Quantum generation of Schwarzschild-de Sitter (Nariai) black holes in effective dilaton-Maxwell gravity, Phys.Lett. B443 (1998) 121, arXiv:hep-th/9808109 
\bibitem{kn:medved}
A.J.M. Medved, Nearly Degenerate dS Horizons from a 2-D Perspective, arXiv:hep-th/0302058 
\bibitem{kn:minic}
Vijay Balasubramanian, Jan de Boer, Djordje Minic, Exploring de Sitter Space and Holography, Class.Quant.Grav. 19 (2002) 5655; Annals Phys. 303 (2003) 59, arXiv:hep-th/0207245 
\bibitem{kn:das}
Sumit R. Das, Thermality in de Sitter and Holography, Phys.Rev. D66 (2002) 025018, arXiv:hep-th/0202008
\bibitem{kn:shank}
S. Shankaranarayanan, Temperature and entropy of Schwarzschild-de Sitter space-time, Phys.Rev. D67 (2003) 084026, arXiv:gr-qc/0301090 
\bibitem{kn:teitelboim}
Andres Gomberoff, Claudio Teitelboim, de Sitter Black Holes with Either of the Two Horizons as a Boundary, Phys.Rev. D67 (2003) 104024, arXiv:hep-th/0302204 
\bibitem{kn:HMS}
Stephen Hawking, Juan Maldacena, Andrew Strominger, DeSitter entropy, quantum entanglement and ADS/CFT, JHEP 0105 (2001) 001, arXiv:hep-th/0002145 
\bibitem{kn:krauss}
Lawrence M. Krauss, Frank Wilczek, Discrete Gauge Symmetry in Continuum Theories, Phys.Rev.Lett. 62 (1989) 1221
\bibitem{kn:preskill}
John Preskill, Lawrence M. Krauss, Local Discrete Symmetry and Quantum Mechanical Hair, Nucl.Phys.B 341 (1990) 50
\bibitem{kn:mcinnes}
Brett McInnes, Alice Universes, Class.Quant.Grav. 14 (1997) 2527


\end{thebibliography}
\end{document}